\documentclass{PoS}

\usepackage{physics}

\title{High-energy neutrino event simulation at NLO in Genie for KM3NeT and other observatories}

\ShortTitle{High-energy neutrino event simulation at NLO in Genie}

\author{The KM3NeT Collaboration$^{\ddagger}$\footnote{for collaboration list see PoS(ICRC2019)1177}\\
{$^{\ddagger}$ \itshape \href{https://www.km3net.org/km3net-author-list-for-icrc-2019}{https://www.km3net.org/km3net-author-list-for-icrc-2019}}\\
E-mail: \email{alfonsog@nikhef.nl}
}

\abstract{High-energy neutrino astronomy is a key pillar of multi-messenger astronomy and has the potential to also advance fundamental neutrino physics. An accurate simulation of the neutrino interactions is key in any analysis of neutrino telescope data.

The currently available generator codes of neutrino interactions use leading order expressions for the differential cross sections which govern rate and kinematics of the events. These calculations are affected by well-known large theoretical uncertainties, hence the need of beyond leading order formalism. Also, a consistent use of modern Parton Density Functions, which are derived using NLO (or NNLO) frameworks, is required.

For this reason and others, the \texttt{GENIE} event generator of neutrino interactions, which is a standard tool in the few-GeV energy regime, has so far been stated to be valid up to only about 1 TeV.

The work reported here consists of a high-energy (up to $10^{10}$ GeV) extension, which will be available in \texttt{GENIEv4}. It interfaces \texttt{PYTHIA6} for the hadronization and employs a pragmatic method to assign the flavour of the struck and outgoing quarks, so that the heavy quark production, including bottom and top, is also correctly simulated. 

\vspace{4mm}
{\bfseries Corresponding authors:}
\speaker{Alfonso Garcia}$^{1}$, Aart Heijboer$^{1}$.\\
{$^{1}$ \itshape Nikhef Institute for Subatomic Physics, Amsterdam, The Netherlands}\\

}

\FullConference{36th International Cosmic Ray Conference -ICRC2019-\\
		July 24th - August 1st, 2019\\
		Madison, WI, U.S.A.}

\begin{document}

\section{Introduction}
\label{sec:intro}

Neutrino telescopes have become a fundamental tool in astroparticle physics: the strong evidence of high energy cosmic neutrinos \cite{ICEEXT}\cite{ANTEXT} and the association of a neutrino event to a flaring blazar \cite{ICEBLAZAR} being key examples. In addition, they can be used to study neutrino properties, such as oscillations \cite{ANTOSC} or cross sections \cite{ICEXSEC}. A new generation of neutrino telescopes is foreseen for the next years with KM3NeT \cite{KM3NET} and IceCube-Gen2 \cite{ICECUBEGEN2}, which will provide unprecedented sensitivity in many analyses. These developments should be accompanied by accurate tools for the simulation of neutrino interactions in or near the detector.

 Monte Carlo simulations of the neutrino interactions rely on computations of the Deep Inelastic Scattering (DIS) neutrino-nucleon cross section. At the moment, leading-order calculations of the double-differential cross section underpin the main neutrino event generator codes (such as GENHEN \cite{GENHEN} (based on LEPTO \cite{LEPTO}), ANIS \cite{ANIS} or NuGen (a modified version of ANIS). In the last years, several groups have provided calculations beyond leading order (i.e NLO or NNLO): the CSMS \cite{Cooper} and BRG \cite{Rojo}.
 
 Such NLO cross-section calculations, however, do not constitute an event generator. A full event generator requires sampling of the full double-differential cross section in an efficient way, and the treatment of the outgoing particles, which result from the breakup of the target nucleon.

 We present a neutrino generator that uses NLO cross sections, and can use the corresponding (also NLO) modern Parton Density Functions (PDFs). The scattering off heavy quarks and the hadronic shower development is simulated by \texttt{PYTHIA6} \cite{PYTHIA6}. The resulting generator, named \texttt{HEDIS}, is implemented as an extension to the \texttt{GENIE} project \cite{GENIE}. \texttt{GENIE} has become a standard tool for long baseline experiments, which detect neutrinos with energies in the 1-10 GeV range. In particular, the (LO) DIS implementation in \texttt{GENIE} is valid only for energies below about 1 TeV. With the inclusion of \texttt{HEDIS} the use of \cite{GENIE} can be extended up to EeV energies. 

In the following sections, we outline the main features of neutrino scattering in the high energy regime and we provide a description of the \texttt{HEDIS} package. Additionally, we present three different calculations of the neutrino cross sections: CSMS, BRG and the calculation used in the KM3NeT LoI \cite{KM3NET}.

\section{Deep Inelastic Scattering}

The neutrino-nucleon scattering can be modelled using the DIS formalism when the transferred momentum is high enough such that the quark content of the target nucleon is resolved. The inclusive cross section of the Charged Current (CC) interaction can be described in terms of the Bjorken scaling variables $x = Q^{2}/2M_{N}$ and $y=(E_{\nu}-E_{lep})/E_{\nu}$ as (neglecting the mass of the lepton)

\begin{equation}
\label{eq:dxsec}
\frac{\dd[2]{\sigma}_{\nu-N}^{CC}}{\dd x \dd y} = \frac{G_{F}^2 M_{N} E_{\nu} M_{W}^{4} }{\pi (Q^2+M_{W}^{2})^2} \left[ y^2x/2F_{1}(x,Q^2) + \left( 1 - y\right)F_{2}(x,Q^2) + y\left( 1 - y^2/2\right)F_{3}(x,Q^2) \right]
\end{equation}

A similar expression can be derived for the Neutral Current (NC) interactions. $F_{1,2,3}$ are the structure functions, which contain all the information about the structure of the nucleon or, in other words, the underlying QCD dynamics.

\label{sec:sf}

The structure functions are the fundamental feature of the DIS model and they can be written as follows (assuming massless partons and the renormalisation and factorization scales both chosen to be $Q^2$)

\begin{equation}
\label{eq:sf}
F_{i}\left( x,Q^2 \right) = \sum_{j} \int_{x}^{1} \frac{\dd z}{z} f_{j} \left(z,Q^2 \right) C_{i,j} \left(x/z,Q^2 \right),
\end{equation}

where $j$ labels gluon and quarks, $f_{j}$ are the PDFs and $C_{i,j}$ are the coefficient functions. 

The PDFs describing the quark and gluon densities in the proton are obtained from fits to hadron collider data, applying the DGLAP formalism \cite{DGLAP}. They form an input to the cross-section calculation and, in GENIE, are obtained via the \texttt{LHAPDF} \cite{LHAPDF} package.  In this program a wide list of different PDF fits is available. They are stored in lookup tables as function of $\log x$ and $\log Q^2$ and some interpolation routines are implemented in order to obtain smooth results. The PDFs are provided in a limited phase space so extrapolation procedures must be used for outer regions.

The coefficient functions that enter into the structure function computation depend on the order of the calculation \cite{COEF}. At LO, most of the coefficients vanish or become delta functions, leading to a linear relation between the structure functions and the PDFs. This makes the structure functions (and thus the differential cross section) fast and easy to compute. Most existing neutrino generators use this approach, relying at the same time on NLO PDFs, despite the inconsistency.

\begin{figure}[h]
\includegraphics[width=1\textwidth]{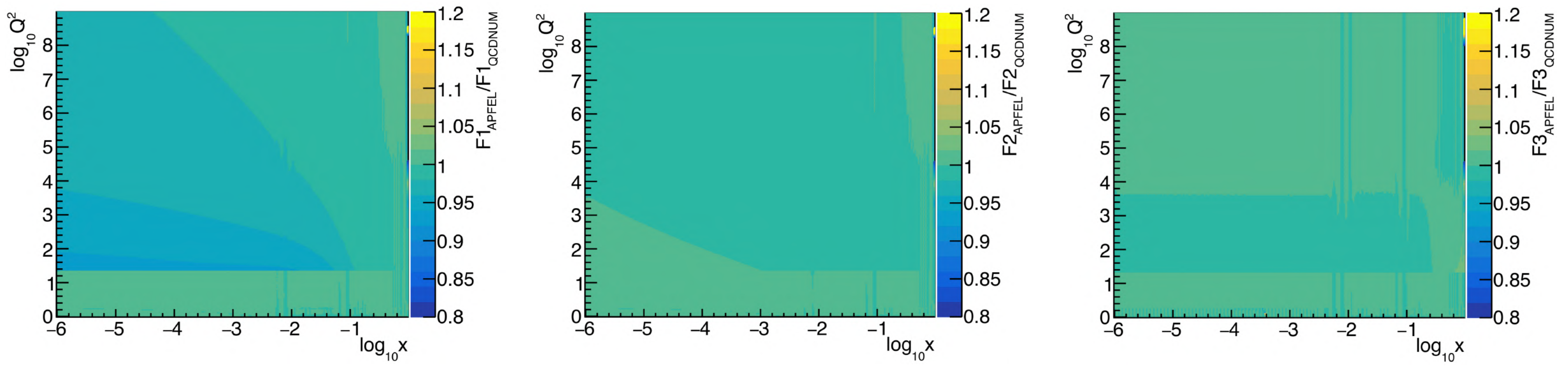}
\caption{Ratio of the \texttt{QCDNUM17} and \texttt{APFEL} proton structure functions for neutrino CC interactions using the HEDIS-CSMS configuration (see Tab.~\ref{tab:conf}). Small discrepancies are found when $Q>m_{charm}$ because the implementation of the heavy-quark mass effects is slightly different in the two codes.}
\label{fig:sf}
\end{figure}

In \texttt{HEDIS}, NLO structure functions are required, which are computed using numerical DGLAP evolution by an external package called \texttt{APFEL} \cite{APFEL}. Another package, called \texttt{QCDNUM} \cite{QCDNUM}, was also tested leading, to similar results as shown in Fig~\ref{fig:sf}. As the calculation of the structure functions is expensive, they are computed in \texttt{HEDIS} prior to the event generation and stored in lookup tables as a function of $\log_{10} x$ and $\log_{10} Q^2$ for each interaction channel. For this, functionality from \texttt{LHAPDF} is used, which has similar tables for the PDFs.

As a result, the application that computes $F_i$ is the central feature of the \texttt{HEDIS} package. Its design allows the manipulation of several features, such as order in pQCD with a desired mass scheme for heavy quark production \cite{Heavy}. As explained in Sec.~\ref{sec:intro}, three different calculations have been tested in this framework. In Tab.~\ref{tab:conf}, their main features are listed.

\begin{table}[h]
\center
\begin{tabular}{ c | c | c | c}
Configuration & HEDIS-CSMS  & HEDIS-BGR & HEDIS-LOI \\
\hline
Order pQCD & NLO  & NLO & LO \\
Mass scheme & ZM-VFNS\footnotemark \cite{ZMVFNS} & FONLL \cite{FONLL} & - \\
PDF & HERAPDF15 \cite{HERA}  & NNPDF31sx+LHCb \cite{NNPDF} & CTEQ6D \cite{CTEQ6} \\
x limits & $[10^{-6},1]$\footnotemark & $[10^{-9},1]$ & $[10^{-6},1]$ \\
$Q^2$ limits & $[1,10^{9}]$ & $[2.69,10^{10}]$ & $[0.1,10^{10}]$ \footnotemark \\
\end{tabular}
\caption{Main features of the three different configurations used in \texttt{HEDIS}. The limits are obtained from the \texttt{LHAPDF6} lookup tables.}
\label{tab:conf}
\end{table}

\subsection{Total cross section}

The total neutrino cross section is important to determine the overall event rate of neutrinos and can be used as a benchmark in the comparison with other published calculations to check the validity of \texttt{HEDIS}.

\footnotetext[1]{In order to match the result from CSMS calculation, the mass of the top quark is not taken into account once 5 flavours are activated.}
\footnotetext[2]{The lower limit is found to be different in \texttt{LHAPDF5} lookup table ($x_{PDFmin}=10^{-8}$).}
\footnotetext[3]{The values of $F_i$ 'freeze' below $Q_{PDFmin}^{2}=1.69$ (i.e. $F_i(x,Q^2) =F_i(x,Q_{PDFmin}^2)$ if $Q^2<Q_{PDFmin}^{2}$).}

The total cross section is computed by integrating the differential expression over the allowed kinematic range in $x$ and $y$. Currently, this integration is performed using a 2 dimensional grid with $\log_{10}$ spacing. More complex algorithms are available, such as VEGAS \cite{VEGAS}, but the performances are found to be very similar. Furthermore, the grid approach allows us to store the maximal differential cross section, which is relevant in the event generation process as explained in Sec.~\ref{sec:evtgen}. Nevertheless, in the future these other integration methods will be available.

The boundaries of the kinematical phase space can be derived using the invarinat mass of the hadronic system $W$ and the four-momentum transfer $Q^2$. Assuming energy and momentum conservation, $M_{N}<W<\sqrt{s}-M_{lep}$ and $E_{lep}-|\vec{p}_{lep}|<(Q^2+M_{lep}^2)/2E_{\nu}<E_{lep}+|\vec{p}_{lep}|$. In practice, the phase space may shrink because the region of validity of the PDFs is smaller (see Sec.~\ref{sec:sf}). Thus, the integration limits depends on the configuration used to compute tables for the structure functions.

At low neutrino energies ($\sim$100 GeV), there is a significant contribution to the cross sections from partons at $Q^2 \equiv 1-3$ GeV. In this range, the PDFs typically have a recommended cut-off as pQCD becomes inapplicable in this regime. As a result large differences between PDFs can be obtained (leading to discrepancies between different calculations if the authors decide to treat the cut-off in different ways). In the future, a more systematic treatment of the low-$Q^2$ regime within \texttt{GENIE} should resolve this.

At the highest neutrino energies, the PDFs are probed at very low $x$-values, which are often poorly constrained by experimental data. 

The total cross section for nuclei is obtained from the appropriate sum of the neutron- and proton cross sections. $F_{i}$ are computed with PDFs from free nucleons, so nuclear-binding effects, such as shadowing \cite{SHADOW}, are not accounted for. None of the configurations presented in this work account for these nuclear effects. However, they could be introduced in future versions by following the procedure described in \cite{Rojo}, where nuclear PDFs are computed using the EPPS16 global analysis \cite{nPDF}.

\begin{figure}
\center
\includegraphics[width=1\textwidth]{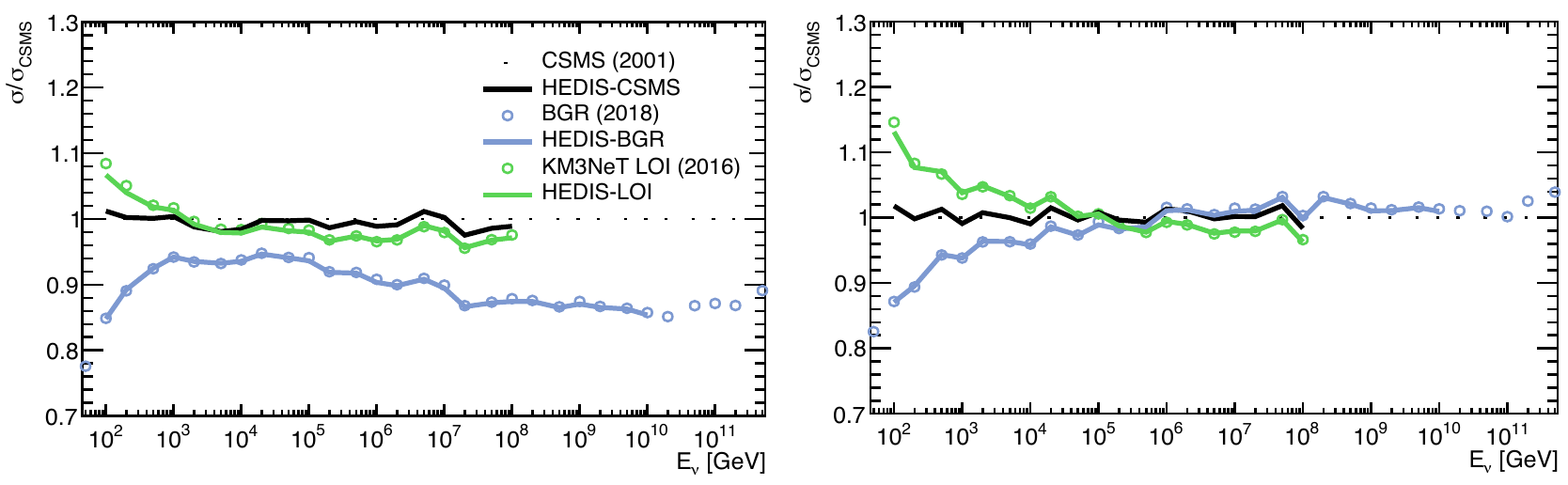}
\caption{Ratio of the total cross section (per nucleon) with respect to CSMS for muon-neutrino scattering with an isoscalar target via CC (left) and NC (right). The results from three different calculations are shown with dots. Lines show the results using \texttt{HEDIS} for the configurations shown in Tab.~\ref{tab:conf}. The predictions from \texttt{HEDIS} are computed up to a certain $E_{\nu}$ because above that value the bulk of the differential cross section lies bellow the $x$ lower limit from the PDFs (obtained from \texttt{LHAPDF6}).}
\label{fig:totxsec}
\end{figure}

Fig.~\ref{fig:totxsec} shows the neutrino total cross section for an isoscalar target using the three different cross-section calculations described in Sec.~\ref{sec:intro}. These predictions have been compared with the results from the \texttt{HEDIS} package using the configurations described in Tab.~\ref{tab:conf}. The result matches at the $1\%$ level in each case, demonstrating that the formalism adopted in \texttt{HEDIS} is accurate. The observed differences between the three calculations at low energies is due to the different $Q^2$ cut-off applied to the PDFs (see Tab.~\ref{tab:conf}). Besides, BRG predicts a lower CC cross section because the production of heavy quarks is more suppressed in the FONLL mass scheme.

\subsection{Hadronization and final state}
\label{sec:hadron}

 In case of a charged-current interaction the outgoing lepton is simulated straightforwardly from the kinematic variables sampled from the double differential cross section. The neutrino interaction will in general result in the breakup of the target nucleon, which produces multiple outgoing particles, which are more complex to simulate.

 The simulation of the hadronic final state is, in GENIE, performed by \texttt{PYTHIA6}, which takes as input a description of the hadronic system just after the $Z^0/W^\pm q \rightarrow  q'$ exchange, in which a quark $q$ from the nucleon has been transformed in a high-momentum quark $q'$. In the input to the \texttt{PYTHIA6} hadronization routine, the state is represented by three particles: one is the outgoing quark produced in by the interaction ($q'$). The other two are baryons or mesons, which are composed by a combination of the quarks from the nucleon remnant and the (sea-quark) companion of the hit parton $q$, $\bar{q}$. 

 The scheme is straightforward to implement at LO, where the flavour of the interacting quark $q$ can be readily sampled from the relative contributions of the PDFs to the cross section, and the flavour of $q'$ follows from this (off-diagonal CKM matrix elements are included). However, beyond leading order, this is not a simple relation, as e.g. gluons will also contribute to the cross section at NLO, but the outgoing quark is not fixed for gluons. As a first pragmatic step, we sample the flavour of the outgoing quark using the LO structure functions. After $x$ and $Q^2$ are sampled from the NLO cross section, the outgoing quark flavour $f$ will be randomly picked with probability:

\begin{equation}
P_f = \sum_j V^{\rm CKM}_{fj} \frac { \sigma^{\rm LO}_f(x,Q^2) }{ \sigma^{\rm LO}_{\rm tot} (x,Q^2) }.
\end{equation}

 In this manner, we benefit from NLO double-differential cross sections, while only using a leading-order approximation in the assignment of the flavour of the outgoing quark. The PDFs used contain heavy flavour quarks as part of the sea, so that also processes like $b \rightarrow t $ are included. For the LO 'flavour-picking', we apply the 'slow rescaling' \cite{Slow} prescription to account for the kinematic suppression of the production of $t,b$ and $c$ quarks. In the NLO cross sections, such effects are accounted for by the mass-scheme.

 We note that the flavour of the outgoing quark is typically of limited relevance in neutrino telescopes as they are not sensitive to microscopic details of the hadronic shower. However, a notable exception is the leptonic decay of heavy flavour, which may produce additional leptons. 
 

\subsection{Event generation}
\label{sec:evtgen}

To simulate neutrino interactions on an event by event basis a sampling method must be used to select the underlying kinematics of the event once the interaction process has been selected. This is usually done using the acceptance-rejection method. In \texttt{HEDIS}, for a particular neutrino energy, a maximal differential cross section is computed and two independent variables ($x$ and $y$) are sampled using the double differential distribution of the chosen process.

\begin{figure}
\center
\includegraphics[width=1.0\textwidth]{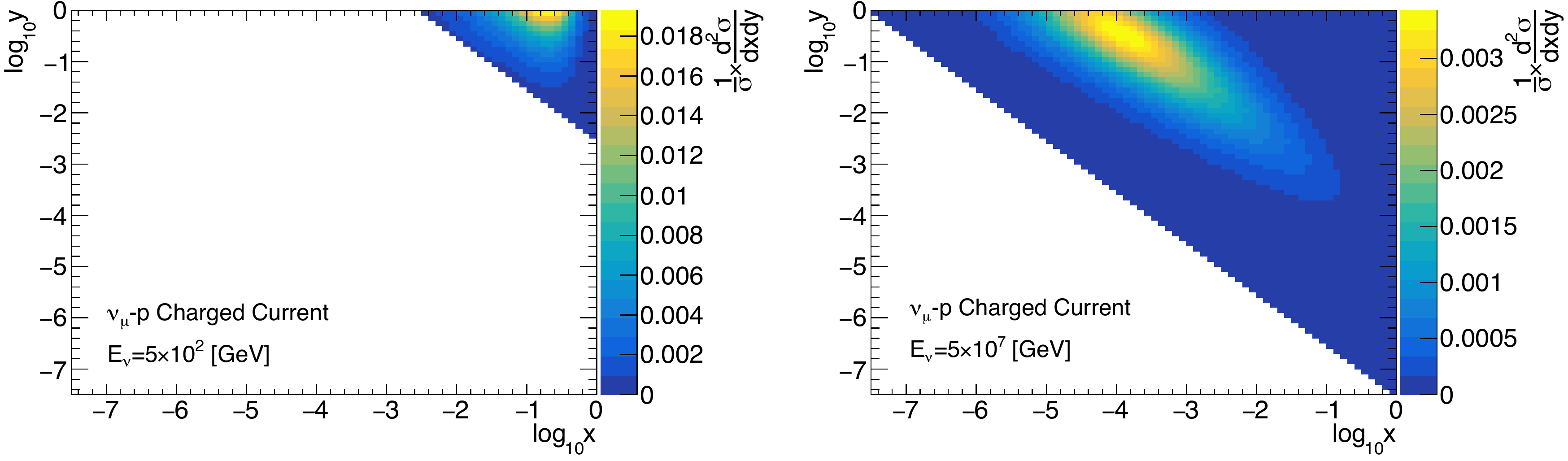}
\caption{Muon-neutrino-proton Charged Current differential cross section for $E_{\nu}=100$ GeV (left) and $E_{\nu}=10^7$ GeV (right). These distributions were obtained using the configuration HEDIS-BGR. The sharp cut-off in the diagonal is due to the $Q^2$ lower limit of the structure functions (see Tab.~\ref{tab:conf}).}
\label{fig:diffxsec}
\end{figure}

Because \texttt{HEDIS} is used in a wide energy range, the phase space of the variables is very broad as shown in Fig~\ref{fig:diffxsec}. Therefore, the sampling is performed over $\log_{10}x$ and $\log_{10}y$ taking into account the corresponding Jacobian. 

\begin{figure}
\center
\includegraphics[width=1.0\textwidth]{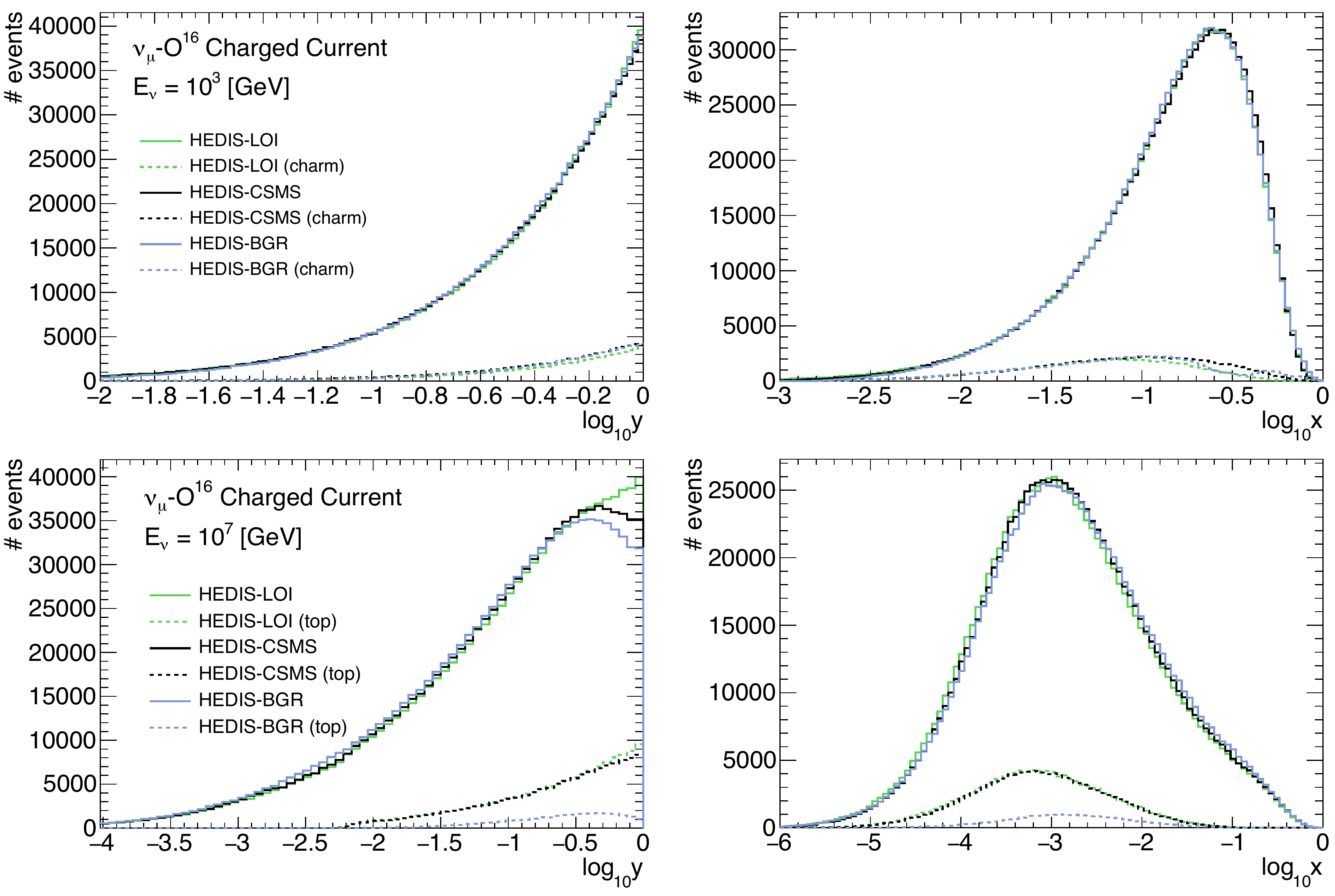}
\caption{Kinematics of muon-netrino-Oxygen Charged Current interactions: $log_{10}y$ (left) and $log_{10}x$ (right). $10^6$ neutrino interactions were simulated using a monochromatic neutrino flux of $E_{\nu}=10^3$ GeV (top) and $E_{\nu}=10^7$ GeV (bottom). Each color represents a configuration described in Tab.~\ref{tab:conf} (nuclear effects are not accounted for). Dashed lines represent the contribution for a particular quark production. }
\label{fig:dist}
\end{figure}

In Fig~\ref{fig:dist}, we compare the $x$ and $y$ distribution for the three configurations (see Tab.~\ref{tab:conf}) using neutrino-Oxygen CC and NC interactions with a monochromatic neutrino flux at two different energies. Overall, the distribution are similar for the three configuration. Nevertheless, the amount of top quarks predicted by BGR differs from the other two configurations. The main reason is that the treatment of the mass for heavy-quark production is more accurate in BGR, leading to a large suppression at these energies. 

\section{Conclusions and Outlook}

The work presented here describes the main feature of the HEDIS package, which will be part of the next release \texttt{GENIEv4}. It will allow to use this Monte Carlo framework to study neutrino interactions at higher energies. The main feature is the new approach to calculate the structure functions, which allows running at higher orders in pQCD. Neutrino observatories will benefit from this extension of the software. Also, it provides theoreticians a tool to test their calculations in an standard way, which will be accessible to experimentalists.  

In addition, we plan to further develop \texttt{HEDIS} in the future. In the short term, it will be important to include in the framework a method to quantify the error of these calculations. Currently, the main uncertainty arrises from the PDFs. Therefore, some extra developments are needed to incorporate them in \texttt{HEDIS}. Furthermore, an extension to \texttt{PYTHIA8} \cite{PYTHIA8} is foreseen for next versions of \texttt{GENIE}, so this new tool will benefit from that. This will allow us to study more sophisticated approaches for the hadronization step, such as parton showering.

More accurate calculations of the Glashow resonance and order sub-leading processes have been developed recently and it would be interesting to include them in this framework \cite{GLASHOW}. Another interesting upgrade is the extension of this model to the low energy regime where most of the long baseline experiments work. In order to do so, calculations with state-of-the-art-PDFs must be studied in the low $Q^2$ regime.
 
\section{Acknowledgements}

We are grateful to Juan Rojo, Rhorry Gauld and Valerio Bertone for their willingness to share their expertise in pQCD. We would like also to thank the Genie Collaboration for supporting this project.


\begin{thebibliography}{99}
\bibitem{ICEEXT} IceCube Collaboration, \emph{Phys. Rev. Lett.} {\bf 113} (2014) 101101 [{\tt 1405.5303}].
\bibitem{ANTEXT} ANTARES Collaboration, \emph{Astrophys. J. Lett} {\bf 853} (2018) 1 [{\tt 1711.07212}].
\bibitem{ICEBLAZAR} IceCube Collaboration, \emph{Science} {\bf 361} (2018) 147 [{\tt 1807.08794}].
\bibitem{ANTOSC} ANTARES Collaboration, \emph{JHEP} {\bf 06} (2019) 113 [{\tt 1812.08650}].
\bibitem{ICEXSEC} IceCube Collaboration, \emph{Nature} {\bf 551} (2017) 596 [{\tt 1711.08119}].
\bibitem{KM3NET} KM3NeT Collaboration, \emph{J. Phys.} {\bf G43} (2016) 084001 [{\tt 1601.07459}].
\bibitem{ICECUBEGEN2} IceCube-Gen2 Collaboration, \emph{PoS} {\bf FRAPWS2016} (2017) 004 [{\tt 1412.5106}].
\bibitem{GENHEN} KM3NeT/ANTARES Collaborations, \emph{Workshop on MC simulation for neutrino telescopes} (2018). 
\bibitem{LEPTO} G. Ingelman et al., \emph{Comput. Phys. Commun.} {\bf 101} (1997) 108 [{\tt hep-ph/9605286}].
\bibitem{ANIS} A. Gazizov et al., \emph{Comput.Phys.Commun.} {\bf 172} (2005) 203 [{\tt astro-ph/0406439}]. 
\bibitem{Cooper} A. Cooper-Sarkar et al., \emph{JHEP} {\bf 08} (2011) 042 [{\tt 1106.3723}].
\bibitem{Rojo} V. Bertone et al., \emph{JHEP} {\bf 01} (2018) 217 [{\tt 1808.02034}].
\bibitem{PYTHIA6} T. Sjostrand et al., \emph{JHEP} {\bf 05} (2006) 2006 [{\tt hep-ph/0603175}].
\bibitem{GENIE} C. Andreopoulos et al., \emph{Nucl. Instrum. Meth.} {\bf A614} (2010) 1 [{\tt astro-ph/0406439}]. 
\bibitem{DGLAP} G. Altarelli et al., \emph{Nucl. Phys.} {\bf B126}  (1977) 298.
\bibitem{LHAPDF} A. Buckley et al., \emph{Eur. Phys. J.} {\bf C75} (2015) 132 [{\tt 1412.7420}].
\bibitem{COEF} A. Vogt et al., \emph{Nucl. Phys. Proc. Suppl.} {\bf 160}  (2006) 44 [{\tt hep-ph/0608307}].
\bibitem{APFEL} V. Bertone et al., \emph{Comput. Phys. Commun.} {\bf 185} (2014) 1647 [{\tt 1310.1394}].
\bibitem{QCDNUM} M. Botje, \emph{Comput. Phys. Commun.} {\bf 182} (2011) 490 [{\tt 1005.1481}].
\bibitem{Heavy} S. Forte et al., \emph{Nucl. Phys.} {\bf B834} (2010) 116 [{\tt 1001.2312}].
\bibitem{ZMVFNS} J. Collins et al., \emph{Nucl. Phys.} {\bf B278} (1986) 934.
\bibitem{FONLL} R. D. Ball et al., \emph{Phys. Lett.} {\bf B754} (2016) 49 [{\tt 1510.00009}].
\bibitem{HERA}  ZEUS, H1 Collaboration, \emph{Proc. 40th Int. Symp. Mult. Dyn.} (2010) [{\tt 1012.1438}].
\bibitem{NNPDF}  R. Gauld et al., \emph{Phys. Rev. Lett.} {\bf 118} (2017) 072001 [{\tt 1610.09373}].
\bibitem{CTEQ6}  J. Pumplin et al., \emph{JHEP} {\bf 07} (2002) 2002 [{\tt hep-ph/0201195}].
\bibitem{VEGAS} T. Ohl, \emph{Comput. Phys. Commun.} {\bf 120} (1999) 13 [{\tt hep-ph/9806432}].
\bibitem{SHADOW} L. Frankfurt, \emph{Nuclei, Phys. Rept.} {\bf 512} (2012) 255 [{\tt 1106.2091}].
\bibitem{nPDF} K. Eskola et al., \emph{Eur. Phys. J.} {\bf C77} (2017) 163 [{\tt 1612.05741}].
\bibitem{Slow} R. M. Barnett, \emph{Phys. Rev. Lett.} {\bf 36} (1976) 1163.
\bibitem{PYTHIA8} T. Sjostrand et al., \emph{Comput. Phys. Commun.} {\bf 178} (2008) 852 [{\tt 0710.3820}].
\bibitem{GLASHOW} R. Gauld, \emph{Nikhef 2019-011} (2019) [{\tt 1905.03792}].




\end{thebibliography}
\end{document}